# Browsing through 3D Representations of Unstructured Picture Collections: an Empirical Study

Olivier Christmann
LORIA, Campus Scientifique
BP 239, Vandœuvre-lès-Nancy Cedex
France
+33 (0)3 83 59 30 83

Olivier.Christmann@loria.fr

Noëlle Carbonell
LORIA, Campus Scientifique
BP 239, Vandœuvre-lès-Nancy Cedex
France
+33 (0)3 83 59 20 32

Noelle.Carbonell@loria.fr

## ABSTRACT
The paper presents a 3D interactive representation of fairly large picture collections which facilitates browsing through unstructured sets of icons or pictures. Implementation of this representation implies choosing between two visualization strategies: users may either manipulate the view (OV) or be immersed in it (IV). The paper first presents this representation, then describes an empirical study (17 participants) aimed at assessing the utility and usability of each view. Subjective judgements in questionnaires and debriefings were varied: 7 participants preferred the IV view, 4 the OV one, and 6 could not choose between the two. Visual acuity and visual exploration strategies seem to have exerted a greater influence on participants' preferences than task performance or feeling of immersion.

## Categories and Subject Descriptors
H.4.3 [**Communications Applications**]: Information browsers
H.5.2 [**User Interfaces**]: Graphical user interfaces.
I.3.6 [**Methodology and Techniques**]: Ergonomics, Interaction techniques.

## 1. CONTEXT AND MOTIVATION
Entertainment and commercial Web-sites, information kiosks and public terminals tend to display a growing number of pictures simultaneously: video and movie stills, CD sleeves, book covers, etc. Personal electronic archives and directories are increasingly cluttered with collections of photographs, scanned documents, videos. It is a standard practice for designers of picture browsers, to display visual information items grouped in 2D arrays that users browse through, using horizontal and vertical scrollbars. Current products (e.g., ACDSee, PhotoSuite or ThumbsPlus) make general use of scrollable 2D arrays of file icons or miniatures for displaying folder contents. Research prototypes of multimedia news summaries [4] or "zoomable" image browsers [1] also use 2D array presentations.

However, browsing through 2D rectangular representations of fairly large collections of 2D icons or pictures (in the form of miniatures) is slow and tiring. In particular, skimming through a whole collection involves two tedious gestural actions which may divert visual attention from the collection, namely, careful horizontal and vertical mouse drags along narrow scrollbars. In addition, searching for a specific item often makes it necessary to go over the whole collection several times, hence to reverse the direction of mouse moves whenever reaching a border of the representation. Such changes of direction may increase users' workload and decrease scanning performances, thus reducing the usability and efficiency of interactive standard 2D representations of graphical objects.

We propose a 3D representation meant to facilitate browsing through fairly large unstructured collections of graphical objects displayed in the form of icons or miniatures. To implement this generic representation, interface designers may choose between two interaction or navigation metaphors:
– direct manipulation of a virtual 3D object representing the whole collection, and interaction with 2D representations of the collection items; or,
– immersion in a virtual 3D environment representing the whole collection, and interaction with this virtual space and the 2D objects that represent the collection items in it.

The empirical study reported here attempts to compare the efficiency and usability of these two interactive visualization paradigms with a view to providing picture browser developers with effective design recommendations.

The next section presents the 3D generic representation we propose for visualizing, and interacting with, collections of graphical items. The following sections describe the empirical study we performed in order to assess the respective efficiency and usability of the two interaction metaphors available to designers for implementing this representation in actual user interfaces.

## 2. GENERIC 3D REPRESENTATION
### 2.1 Description
The proposed representation seems most appropriate for browsing collections including a number of graphical objects in the region of one thousand, on standard PC displays (19'' or 21'' screen size). Graphical objects in collections of this size should be in the form of icons or miniatures.

For any such collection, our proposal is to arrange graphical items in rows and columns and to plaster them on the inner or outer wall of a vertical cylinder; see figure 1. Rotation of the cylinder to the left or to the right is activated by means of two arrows at the bottom of the display. Therefore, repeated surveys of the entire collection are performed without changing the direction of the cylinder motion. Clicking on one arrow launches the rotation; clicking on the right button of the mouse stops it. Rotation speed can be controlled by means of the mouse scroll wheel. Two other arrows are available for adjusting the apparent distance between the user and the virtual cylinder to their visual capabilities. One sixth of the cylinder lateral surface only is displayed at once, in order to reduce the negative effects of perspective distortions on visual search without loosing the useful focusing effects of such distortions, namely, enlargement of pictures in the lateral columns for the inner view of the cylinder, and in the central column(s) for the outer view. Standard zooming facilities (fixed size of the zoomed image) are available for enlarging icons or miniatures.

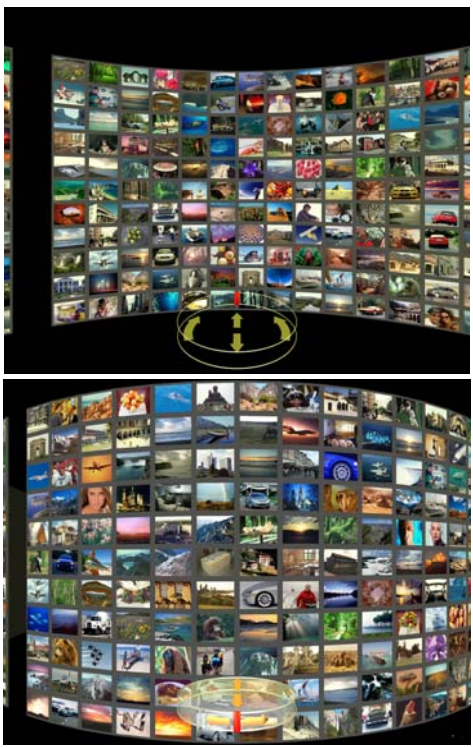

**Figure 1. Inner and outer views (same number of photos).**

A vertical "slit" in the cylinder (i.e., an empty column) helps users to keep track of navigation progress; they may use it as a full turn completion indicator. Additionally, at the bottom of the display, a 3D miniature cylinder with a vertical slit reproduces the motion of the representation of the collection. The current position of the column displayed at the centre of the screen is materialized on the miniature cylinder by a red vertical line. The motions of both cylinders are synchronised, thus providing users with continuous navigation feedback.

### 2.2 Comments
The inner and outer cylindrical representations of a collection are not equivalent; they differ from each other in their visual properties and their possible influence on users' subjective judgements. See figure 1. Firstly, for the same number of displayed rows and columns, the size of items in the inner view increases from the centre of the display towards its right and left boundaries, while it decreases from the centre towards side boundaries in the outer view. In addition, enlarged items are less distorted in the inner view than in the outer view. These differences may influence users' visual exploration strategies. Secondly, the convexity of the outer view may induce users to view interaction with this view as manipulation of virtual objects, while the concavity of the inner view will make them feel immersed in the virtual representation space. This difference in system image is likely to influence subjective judgements.

The diversity and scientific interest of the ergonomic issues stemming from these differences are the main research motivations underlying the study presented in the following sections. The simplicity of the proposed generic representation and its originality also count among our motivations. Design simplicity facilitates software implementation and ergonomic evaluation. As for simplicity of usage, it is a major asset for ensuring the success of a software product intended for the general public. Originality here refers to the absence of published research on this representation. Up to now, research on visualization techniques has mainly focused on interactive 2D representations of very large structured data sets, such as hyperbolic trees or treemaps [5]. In addition, the use of such techniques for visualizing collections of graphical objects has motivated but a few studies, such as [1] which reports the use of treemaps for presenting photographs. A few 3D visualization techniques only have been proposed. Cone trees [2] are meant for representing structured data while Card's 3D wall [3] can be used for presenting unstructured and structured sets of textual, digital or graphical items. It is composed of a front wall parallel to the screen and flanked by two side walls. This 3D representation has obvious similarities with the outer view of the representation we propose. However, it is devoid of the advantages provided by a cylindrical representation with a current position indicator.

## 3. EVALUATION PROCEDURE
### 3.1 Design and implementation of both views
The representation described in the preceding section has been implemented as follows. The chosen application domain was photograph browsing. We used colour photographs pulled from popular Web sites. Collections included 869 photographs, arranged along 11 rows and 79 columns (80 with the empty column). 14 columns or so were visible at once. Maximum rotation speed was 7 times the default minimum speed. The size of zoomed images was 300 x 225 pixels on a screen with a 1280 x 1024 pixel resolution. Computer displays were projected on a wall screen (size: 1.00 x 0.75 m; distance between participant and screen: 2.20 m). Interaction facilities with the inner (IV) and outer (OV) views were identical; see figure 1. The virtual reality environment AReVi was used for all software developments.

### 3.2 Tasks

Participants performed two types of visual search tasks which are frequently used while browsing through collections of pictures:
- T1: looking for pictures matching a set of criteria; these pictures and their locations in the collection are unknown.
- T2: looking for a visually familiar picture; the location of the picture in the collection may be known or unknown.

Participants performed 10 tasks of each type for each of the two views. All T1 tasks (20) were carried out first, and processing order was counterbalanced between participants as follows:
- T1[IV then OV], T2[IV then OV] for half the participants,
- T1[OV then IV], T2[OV then IV] for the other half.

Each T1 task series was preceded by a short training (2 tasks).

To achieve a T1 task scenario, participants had to search for and select a unique target photograph in the currently displayed collection, using an unambiguous written description of the target's topic, spatial layout, dominant colours and specific details. The structure and information content of target descriptions were standardized in order to homogenize task difficulty and help users to develop efficient reading and searching routines. Following target selection, the system demonstrated the correct solution, so that all participants had the opportunity to watch the target at least once.

Each T2 task consisted in first displaying one of the targets used for T1 at the centre of the screen for three seconds. Then, participants started searching for it on the 3D cylinder and, when they had found it, selected it. T1 targets were visually familiar to participants since they had already seen them once or twice during T1. Many participants remembered the locations of some targets more or less precisely, especially the row they belonged to. Search time was bounded (3 min.) for both tasks.

### 3.3 Participants

17 experienced computer users, 7 female and 10 male graduate or PhD students, participated in the experiment. The number of participants being rather small, we chose a narrow age range (from 21 to 30 years) in order to limit the heterogeneity of the group regarding task execution performance, interests and preferences. All participants had normal sight according to the Bioptor test kit (Stereo Optical Company, Inc.

### 3.4 Data collection

Participants' interactions with both views were logged up. In addition, after the experimentation, all participants filled in a questionnaire meant to elicit their reactions, subjective judgements, and comments, and each of them participated in a debriefing interview.

## 4. FIRST RESULTS

Participants' preferences and performances are detailed in table 1. Preferences have been established from questionnaires and debriefing transcripts. Performance measures, computed from participants' logs for T1 and T2 tasks, include:
- the number of search failures for each view, IV and OV,
- the difference between IV and OV search times, and
- the number of cylinder rotations for each view.

The order in which participants interacted with the two views had no significant influence on their performances. Due to a technical incident, data from participant 8 have been ignored. Some participants expressed judgements during the debriefings that differed slightly from those in questionnaires; both judgements have been reported in table 1.

**Table 1. Participants' preferences and performances.**

Part.: Participant numbers.
Pref.: Preferences from questionnaires (Q) and debriefings (D) for the inner (I) or outer (O) view. "<" denotes a weak preference, and "H" hesitation between the two views.
Diff. Err.: Differences* in error numbers between IV and OV.
Diff. T: Differences* in cumulated task completion times between IV and OV (percentages being computed over the shortest time).
Diff. Rot.: Differences* in numbers of full rotations between IV and OV.

* i.e., OV_value – IV_value.

| Part. | Pref. | | Diff. Err. | | Diff. T. % | | Diff.. Rot. | |
|---|---|---|---|---|---|---|---|---|
| | Q | D | T1 | T2 | T1 | T2 | T1 | T2 |
| 1 | <I | - | 1 | - | 38 | 13 | 0 | - |
| 2 | <O | H | -1 | 0 | -55 | 39 | 1 | 3 |
| 3 | H | I | 4 | -1 | 33 | -29 | 2 | -5 |
| 4 | H | O | 1 | 0 | -5 | 32 | 0 | 1 |
| 5 | <I | I | 0 | 0 | 19 | -41 | 0 | -1 |
| 6 | <<I | <I | -1 | 1 | -5 | 7 | 2 | 3 |
| 7 | O | - | 3 | -1 | 57 | -60 | 1 | -7 |
| 9 | O | - | 0 | 0 | -19 | -9 | 0 | -4 |
| 10 | O | - | 0 | 0 | -11 | -26 | -3 | -3 |
| 11 | I | - | 1 | 2 | -11 | 147 | -3 | 1 |
| 12 | H | - | 1 | 0 | 74 | 66 | 0 | 1 |
| 13 | H | - | 1 | 0 | 22 | -29 | -4 | -6 |
| 14 | <I | I | 3 | 0 | 83 | 52 | 0 | 2 |
| 15 | <I | I | 2 | 0 | 51 | 6 | 2 | -3 |
| 16 | O | - | 0 | 0 | -23 | 78 | -8 | 2 |
| 17 | <I | - | -1 | -2 | -51 | -15 | -5 | -3 |
| 18 | <O | H | 2 | 0 | 104 | 12 | 1 | -1 |

According to the debriefings, all participants save one judged both 3D views positively, and preferred them to 2D representations. Contrastingly, their attitudes towards each of the 3D views disagreed. Seven participants preferred the inner view, four favoured the outer view, and six could not choose between the two views; the last group includes participants who expressed different preferences in the questionnaire and during the debriefing. A first tentative conclusion can be drawn from the diversity of participants' preferences: in order to meet with general user acceptance, the inner and outer views of the proposed representation should be both implemented in picture browsers.

Participants' performances, questionnaires and debriefings were further analysed so as to gain an insight into the factors that may have influenced their subjective judgements. To refine the three user profiles detected, we analysed the results of the vision tests, and elicited participants' visual exploration strategies using debriefings and interaction logs (especially the positions of zoomed pictures on the display). Our aim was to determine whether participants' preferences had been influenced by the specific visual properties of each view, mainly the location of enlarged[1] pictures which varies from one view to the other (see subsection 2.2). Participants were found to use two main visual exploration strategies: they focused their gaze either on incoming columns (In strategy) or on the central column(s) of the display (C strategy). Synthesized data are presented in table 2, grouped by user profile; GI designates the group of participants who preferred the inner view, GO the participants who liked the outer view better, and GH those who hesitated between the two views.

**Table 2. Performances and visual exploration strategies of participants grouped by preferences, GI, GO, GH.**

Vis. Ac.: Visual acuity; normal (N), low (L).
I, O: inner (O) and outer (I) views.
Err., T., Rot.: View (I or O) for which participants achieved best results regarding error number (Err.), task completion time (T), number of full rotations (Rot.), respectively.
Strat. (T1): Visual exploration strategies for T1 tasks.
In, C: Observation of the display focused on incoming/central column(s).

**Table 2.1. GI group**

| Participant | | 1 | 5 | 6 | 11 | 14 | 15 | 17 |
|---|---|---|---|---|---|---|---|---|
| Vis. Ac. | | N | N | L | N | N | N | L |
| Err. | T1 | - | - | - | - | I | I | - |
| | T2 | - | - | - | I | - | - | O |
| T. | T1 | I | I | - | O | I | I | O |
| | T2 | I | O | - | I | I | - | O |
| Rot. | T1 | - | - | I | O | - | I | O |
| | T2 | - | - | I | - | I | O | O |
| Strat. (T1) | IV | In | In | In | In | C | C | In |
| | OV | In | In | In | In | C | C | In |

**Table 2.2. GO group and GH group**

| Participant | | GO group | | | | GH group | | | | | |
|---|---|---|---|---|---|---|---|---|---|---|---|
| | | 7 | 9 | 10 | 16 | 2 | 3 | 4 | 12 | 13 | 18 |
| Vis. Ac. | | N | L | L | L | N | N | L | N | N | N |
| Err. | T1 | I | - | - | - | - | I | - | - | - | I |
| | T2 | - | - | - | - | - | - | - | - | - | - |
| T. | T1 | I | O | O | O | O | I | - | I | I | I |
| | T2 | O | - | O | I | I | O | I | I | O | I |
| Rot. | T1 | - | - | O | O | - | I | - | - | O | - |
| | T2 | O | O | O | I | I | O | - | I | O | - |
| Strat. (T1) | IV | ? | ? | In | C | C | In | ? | ? | In | In |
| | OV | C | C | In | In | ? | C | ? | ? | C | In |

Task performance may have influenced the attitudes of some participants. However, visual acuity and exploration strategies are factors which seem to have exerted a greater influence on preferences. 5 participants (out of 7) in the GI group had normal visual acuity while 3 (out of 4) participants in the GO group had lower visual acuity. Similarly, 5 GI participants used the In strategy consistently versus only one in the GO group.

---
[1] Due to 3D perspective rules.

Contrastingly, performances in both groups vary from one type of tasks to the other, and from one criterion to another. Participant 17 in the GI group even achieved better performances using the OV view with respect to all criteria. As expected, data from participants in the GH group show a much greater diversity.
Interestingly, the immersive feature of the IV view seems to have exerted a limited influence on judgements. Four participants only alluded to it during the debriefings: S9 in the GO group mentioned it as a negative feature, saying that she felt ill at ease when using the IV view, while S5 and S11 in the GI group and S18 in the GH group referred to it as an attractive feature.
Designers of adaptable or adaptive browsers of graphical objects could take advantage of these findings. The effectiveness of static and dynamic user models could be greatly improved by taking these factors into account in their design.

# 5. CONCLUSION

We have presented an empirical study meant to assess the utility and usability of a generic cylindrical 3D representation of fairly large unstructured sets of graphical objects. This representation is meant to facilitate browsing. Two metaphors are available for designing interaction with this representation: users may either manipulate the cylinder (outer view) or feel immersed in it (inner view). 17 participants interacted with both views successively. 7 participants preferred the inner view, 4 the outer one, and 6 could not decide between the two. Analysis of performances, satisfaction questionnaires and debriefings suggests that participants' preferences were less influenced by task performance and subjective perceptual 3D effects than by visual acuity and visual exploration strategy. Participants who had normal visual acuity and focused their visual attention on incoming columns seem to have been biased towards the inner view, while participants who had lower visual acuity and focused their gaze on columns at the centre of the display may have been biased towards the outer view. 4 participants only mentioned the feeling of immersion induced by the inner view to explain their subjective judgements; the specific visual characteristics of each view, especially picture size and distortion, location of enlarged pictures (in the centre or on the sides of the display) were alluded to much more often for justifying preferences.

We are currently preparing an experimental study meant to compare the usability of both 3D views with standard 2D displays for carrying out T1 and T2 type tasks. As visual exploration strategy seems to have a significant influence on user preferences, the eye movements of a few participants will be recorded and analysed.

# 6. REFERENCES

[1] Bederson, B.B. (2001). PhotoMesa: a Zoomable Image Browser Using Quantum Treemaps and Bubblemaps. Proc. UIST'01, CHI Letters, 3(2), 71-80.

[2] Hearst, M., Karadi, C. (1997). Cat-a-cone: an interactive interface for specifying searches and viewing retrieval results using large category hierarchy. Proc. SIGIR, pp. 246-255.

{3] Mackinlay, J.D., Robertson, G.G., Card, S.K. (1991). Perspective Wall: Detail and Context Smoothly Integrated. Proc. CHI'91, ACM Press, New York pp. 173-179.

[4] Maybury, M. (2000). News on demand: Introduction. Communications of the ACM, 43(2), 32-34.

[5] Pirolli, P., Card, S.K., Van Der Wege, M. (2000). The Effect of Information Scent on Searching Information Visualizations of Large Tree Structures. Proc. ACM AVI'00, pp. 161-172.